\documentclass[twocolumn,superscriptaddress,showpacs,aps,amsmath,amssymb,prb]{revtex4-1}
\usepackage{amsfonts}
\usepackage{mathrsfs}

\usepackage{bm}
\usepackage{graphicx}

\newcommand{\ti}{\Tilde}
\newcommand{\nl}{\nonumber \\}

\newcommand{\up}{\uparrow}
\newcommand{\down}{\downarrow}
\newcommand{\leftact}{\overset{\rightarrow}}
\newcommand{\rightact}{\overset{\leftarrow}}

\newcommand{\Sec}[1]{Sec.\;\ref{#1}}

\newcommand{\be}{\begin{equation}}
\newcommand{\ee}{\end{equation}}
\newcommand{\bea}{\begin{eqnarray}}
\newcommand{\eea}{\end{eqnarray}}
\newcommand{\bal}{\begin{align}}
\newcommand{\eal}{\end{align}}
\newcommand{\bsube}{\begin{subequations}}
\newcommand{\esube}{\end{subequations}}
\newcommand{\Eq}[1]{Eq.\,(\ref{#1})}
\newcommand{\Eqs}[1]{Eqs.\,(\ref{#1})}
\newcommand{\Fig}[1]{Fig.\,\ref{#1}}
\newcommand{\dg}{\dagger}
\newcommand{\la}{\langle}
\newcommand{\ra}{\rangle}

\begin{document}

\title{
  Non-Markovian shot noise spectrum of quantum transport through quantum dots
}

\author{Jinshuang Jin}\email{hznu.jin@gmail.com}
\affiliation{ Department of Physics, Hangzhou Normal University,
  Hangzhou 310036, China}
\affiliation{Department of Chemistry, Hong
Kong University
   of Science and Technology, Kowloon, Hong Kong}

\author{Xin-Qi Li}
\affiliation{Department of Chemistry, Hong Kong University
   of Science and Technology, Kowloon, Hong Kong}
\affiliation{ Department of Physics, Beijing Normal University,
Beijing 100875, China}

\author{Meng Luo}
\author{YiJing Yan}\email{yyan@ust.hk}
\affiliation{Department of Chemistry, Hong Kong University
   of Science and Technology, Kowloon, Hong Kong}

\date{\today}

\begin{abstract}
 The generalized quantum master equation with
transport particle number resolution,
like its conventional unconditioned counterpart,
has also the time--local and time--nonlocal prescriptions.
The latter is found to be more suitable
for the effect of electrodes bandwidth
on quantum transport and noise spectrum
for weak system--reservoir coupling,
as calibrated with the exact results in the absence of Coulomb interaction.
We further analyze the effect of Coulomb interaction on
the noise spectrum of transport current through
quantum dot systems, and
show that the realistic
finite Coulomb interaction and finite bandwidth are manifested only
with non-Markovian treatment.
We demonstrate a number of non--Markovian characteristics
of shot noise spectrum, including that due to finite bandwidth
and that sensitive to and enhanced by the magnitude of Coulomb
interaction.

\end{abstract}

\maketitle

\section{Introduction}
\label{intro}
$\mathscr{F}$ $\cal F$
Shot noise due to charge discreteness in mesoscopic transport has
stimulated great interest in recent years. It provides additional
information beyond the average current, especially on the nature of
fluctuating environment coupling to the mesocopic system. \cite{Bla001,Naz03}
Conventionally, evaluations of shot noise and
higher cumulants of current in full
counting statistics are largely restricted to zero frequency, and
Born--Markov master equation approach is employed.
\cite{Sun9910748,Fli04205334,Kie06033312,Wan07125416}
Memory effects of fluctuating environment on the first few
cumulants of current at zero frequency were investigated recently,
revealing that non-Markovian corrections are increasingly
important to higher cumulants. \cite{Bra06026805,Fli08150601}
The related features are even more pronounced at high frequency,
as demonstrated experimentally. \cite{Agu001986,Zak07236803,Ona06176601}
   Non-Markovian feature manifests itself the
nature of fluctuating environment.
Flindt {\it et al} \cite{Fli08150601} and Aguado {\it et al} \cite{Agu04206601}
studied the noise spectrum of qubit under transport,
with non-Markovian treatment of the phonon bath environment,
but considered electrodes (electron reservoirs)
in the Markovian and large voltage limit.
Non-Markovian characteristics of electron reservoirs differ
distinctly from that due to bosonic--bath coupling.
Their effects on the frequency--resolved shot noise
have been explored in the wide--band limit (WBL),
\cite{Eng04136602,Ent07193308,Rot09075307}
and the appearance of step structure reflects directly
the discreteness of energy
levels of the dots.

 In this work, we demonstrate some basic
non-Markovian features of shot noise, resulted from the
finite bandwidth property of electrodes
and the finite
Coulomb interaction of mesoscopic systems.
The present calculation is based on the
particle--number resolved or generalized
quantum master equation (GQME),
together with MacDonald's formula.
Like its conventional unconditioned counterpart,
there are two prescriptions, i.e.,
the time--local (TL) versus time--nonlocal (TNL)
forms of GQME, and they are not equivalent
in the weak system--environment interaction treatment.
For phonon bath environment such as spin--boson system
and optical line shape problems,
it often found that the TL ansatz
is superior.\cite{Yan05187,Che09094502}
 For electrons reservoirs
environment for quantum transport,
however, the TNL prescription rather
is more appropriate.
The resulted expression of the noise spectrum contains
explicitly the memory
effects due to finite electrodes bandwidth.
The superiority of TNL--GQME over
TL--GQME\cite{Leh02228305,Li05205304,Li05066803,Li07075114,Har06235309}
will be verified by comparison with
an exact path-integral theory\cite{Jin10083013,Jin08234703,Zhe08184112,Zhe08093016}
in the absence of Coulomb
interaction.

 The paper is organized as follows.
In \Sec{ththeo1}, we present the TNL--GQME,
viewed from both the particle number aspect
and its conjugated counting field aspect,
for full counting statistics.
The resulting transport current noise spectrum
formalism is given in \Sec{ththeo2},
together with general
remarks on non-Markovian shot noise characteristics.
In \Sec{num}, we implement the proposed scheme to some
noninteracting and interacting model quantum dots.
Including in the noninteracting case
is also the exact result that justifies
the present TNL--GQME approach, while
discriminates its TL counterpart.
Finally, we conclude in \Sec{sum}.

\section{Generalized quantum master equation approach}
\label{ththeo1}

\subsection{Decomposition of conventional memory kernel}
\label{ththeo1A}

 It is noticed that the GQME for full counting statistics
can be constructed rather straightforwardly
by using the counting field-dressed
method;\cite{Lev93225,Lev964845,Bag03085316}
cf.\ \Eq{ME-1} and comments there.
Here, we like to provide also an alternative view
that may shed light on how the counting measurement field
selects the transport components from the total
dissipation superoperator, denoted below as $\hat\Sigma(t)$,
in the conventional or
counting field-free QME theory [cf.\ \Eq{ME0}].
In the present weak coupling
theory the total dissipation superoperator
is additive, i.e., $\hat\Sigma(t)=\hat\Sigma^{\rm el}(t)+
\hat\Sigma^{\rm ph}(t)$,
for its contributions
from electron reservoirs and phonon bath interactions.
While $\hat\Sigma^{\rm el}(t)$
contains both transport and non-transport components,
the phonon-bath induced $\hat\Sigma^{\rm ph}(t)$ is
itself non-transport,
but destroys the coherence in central system.

  The conventional QME in memory kernel
prescription for the reduced system density operator reads\cite{Yan05187}
\be\label{ME0}
 \dot\rho(t) = -i{\cal L}\rho(t)
  - \int_{-\infty}^t\!\!{\mathrm d}\tau
  \hat \Sigma(t-\tau)\rho(\tau).
\ee
Here, $\mathcal{L}\,\cdot\equiv [H,\cdot\,]$ is the reduced
quantum dots system Liouvillian;
$\hat \Sigma(t)$ denotes the dissipation kernel
superoperator for the coupling environment effect
on the reduced transport system.
Assume the weak system-environment coupling.
It leads to
 $\hat \Sigma(t-\tau)
=\la {\cal L}'(t)e^{-i\mathcal{L}(t-\tau)}
 {\cal L}'(\tau)\ra_{\rm env}$, with
$\mathcal{L}'(t)\,\cdot\equiv[H'(t),\cdot\,]$
being the system--environment coupling Liouvillian
and $\la \cdots\ra_{\rm env}$ denoting the average over
environment degrees of freedom, including
both electron reservoirs and phonon bath.
Throughout this work, we set the Planck constant
and electron charge $\hbar=e=1$.

  For clarify, let us treat explicitly
only the influence of electron reservoirs
of coupling
electrodes ($\alpha= \textrm{L and R}$).
They are modeled by noninteracting electrons,
 $h_\textrm{res}=\sum_{\alpha} h_{\alpha}
=\sum_{\alpha k} \epsilon_{\alpha k}
  c^{\dg}_{\alpha k}c_{\alpha k}$,
and their coupling with system responsible for
transport current is
\be\label{Hprime0}
  H'_{\text{sys-res}}=\sum_{\alpha k\mu} (t_{\alpha k\mu}
   c^{\dg}_{\alpha k}d_{\mu }+t^{\ast}_{\alpha k\mu}
   d^{\dg}_{\mu }c_{\alpha k}).
\ee
Here, $d_{\mu }$ ($d^{\dg}_{\mu}$)
is the annihilation (creation) operator for an electron in
the specified spin-orbital level of the quantum dots system,
while
$c_{\alpha k}$ ($c^{\dg}_{\alpha k}$)
is that of the specified $\alpha$--electrode level with
energy $\epsilon_{\alpha k}$.
 In the $h_\textrm{res}$--interaction picture,
\Eq{Hprime0} is
\be\label{Hprime}
  H'_{\text{sys-res}}(t)
 =\sum_{\alpha\mu}\big[F^{(+)}_{\alpha\mu}(t)d_{\mu}
 +d^{\dg}_{\mu}F^{(-)}_{\alpha\mu}(t)\big].
\ee
where $F^{(+)}_{\alpha\mu}(t)\equiv
\sum_k e^{ih_{\alpha}t} \big(t_{\alpha k\mu}
 c^{\dg}_{\alpha k}\big)e^{-ih_{\alpha}t}
 \equiv [F^{(-)}_{\alpha\mu}(t)]^{\dg}$
are the stochastic reservoir operators.
They satisfy
the Gaussian statistics with Wick¡¯s theorem for thermodynamic
average. As a result, the effects of
reservoirs on the reduced system can be completely determined
by the reservoir correlation functions,
\be\label{Ct}
  C^{(\pm)}_{\alpha\mu\nu}(t-\tau)
 = \la F^{(\pm)}_{\alpha\mu}(t) F^{(\mp)}_{\alpha \nu}(\tau)
  \ra_\textrm{\rm res}.
\ee
For bookkeeping in the following, we denote $\sigma=+$ or $-$,
and $\bar\sigma$ be the opposite sign of $\sigma$.
Denote also
$d^{+}_{\mu}\equiv d^{\dg}_{\mu}$ and $d^{-}_{\mu}\equiv d_{\mu}$.

Treating $H'_{\text{sys-res}}(t)$ up to second order, the convolution term
in \Eq{ME0} is explicitly expressed
as \cite{Yan982721,Yan05187,Mak01357,Shn0211618}
\begin{align}\label{memoryt}
 \hat\Sigma(t)\!\otimes\!\rho(t)
&\equiv \int_{-\infty}^t\!\!{\mathrm d}\tau
  \hat \Sigma(t-\tau)\rho(\tau)
\nl&=
 \sum_{\sigma\alpha\mu\nu}\Big\{
  \big[d^{\bar\sigma}_\mu,
     \big(C^{(\sigma)}_{\alpha\mu\nu}(t)e^{-i{\cal L}t}\big)
   \!\otimes\! \big(d^{\sigma}_\nu\rho(t)\big)\big]
\nl &\quad\ \ \
 +\big[\big(C^{(\sigma)\ast}_{\alpha\mu\nu}(t)e^{-i{\cal L}t}\big)
   \!\otimes\! \big(\rho(t)d^{\bar\sigma}_\nu\big), d^{\sigma}_\mu\big]
\Big\} .
\end{align}
Let $\mathscr{L}[x(t)]$ be the Laplace frequency transformation
of an arbitrary function of time $x(t)$; e.g.,
\be \label{LapTran}
  C^{(\sigma)}_{\alpha\mu\nu}(\omega)\equiv
    \mathscr{L}[C^{(\sigma)}_{\alpha\mu\nu}(t)]
   \equiv \int^\infty_0 \!\!{\mathrm d}t\,
   e^{i\omega t}C^{(\sigma)}_{\alpha\mu\nu}(t).
\ee
The Liouville--space self--energy is
$\Sigma(\omega) \equiv \mathscr{L}[\hat\Sigma(t)]$.
Together with the notion of
$\leftact{d^{\sigma}_{\mu}}\hat O \equiv {d^{\sigma}_{\mu}}\hat O$ and
$\rightact{d^{\sigma}_{\mu}}\hat O \equiv \hat O {d^{\sigma}_{\mu}}$,
we recast self--energy in \Eq{memoryt} as
\begin{align}\label{memoryw}
\Sigma(\omega)=\Sigma^{(0)}(\omega)
-\sum_{\alpha}\left[\Sigma^{(+)}_{\alpha}(\omega)+\Sigma^{(-)}_{\alpha}(\omega)\right],
\end{align}
with
\bsube\label{allSig}
\be\label{Sig0}
 \Sigma^{(0)}(\omega)= \!\sum_{\sigma\alpha\mu\nu}\!
   \left[\leftact{d^{\bar\sigma}_{\mu}}
     C^{(\sigma)}_{\alpha\mu\nu}(\omega\!-\!{\cal L})
   \leftact{d^{\sigma}_{\nu}}
  +
   \rightact{d^{\sigma}_{\mu}}
     C^{(\sigma)\ast}_{\alpha\mu\nu}({\cal L}\!-\!\omega)
   \rightact{d^{\bar\sigma}_{\nu}}\right],
\ee
and
\be\label{Sigpm}
 \Sigma^{(\sigma)}_{\alpha}(\omega)= \!\sum_{\mu\nu}\!
   \left[\rightact{d^{\bar\sigma}_{\mu}}
    C^{(\sigma)}_{\alpha\mu\nu}\!(\omega\!-\!{\cal L})
   \leftact{d^{\sigma}_{\nu}}
  +  \leftact{d^{\sigma}_{\mu}}
       C^{(\sigma)\ast}_{\alpha\mu\nu}\!({\cal L}\!-\!\omega)
   \rightact{d^{\bar\sigma}_{\nu}}\right].
\ee
\esube
The kernel of $\Sigma^{(0)}(\omega)$ does not change the electron particle number,
and it contains in general also
the phonon bath component,
$\Sigma^{\rm ph}(\omega)$, as discussed earlier.
On the other hand, $\Sigma^{(\sigma)}_{\alpha}(\omega)$
associates with increase ($\sigma=+$)
or decrease ($\sigma=-$) of particle number
by one. The corresponding $\Sigma^{(\pm)}_{\alpha}(t)$
defines the transport memory kernel.
The above picture is closely related to the counting
statistics to be elaborated in the coming two
subsections.

\subsection{Generalized quantum master equation for counting statistics}
\label{ththeo1B}
  Rather than the above conventional QME (\ref{ME0})
for the \emph{unconditional} $\rho(t)$, a richer
information contained equation for \emph{conditional state} will be
more desirable. This is the GQME
for {\it particle-number-resolved} $\rho^{(n)}(t)$,%
\cite{Naz03,Li05205304}
the reduced state {\it conditioned} by the given number
$n$ of electrons transmitted, within the measuring {\em time internal} $t$,
through the specified lead that will
be denoted {\em implicitly} as the electrode $\alpha$ hereafter.
While the unconditional state is
$\rho(t)=\sum_n\rho^{(n)}(t)$,
the conditional one is related to the
current counting distribution function,
$P(n,t)\equiv {\rm Tr}\big[\rho^{(n)}(t)\big]$,
which contains full information
including current, shot noise, and
all higher moments of current fluctuations. \cite{Li05205304}
The GQME with transmitted particle
number resolution describes the
quantum evolution in relation to
the distribution function $P(n,t)$.

 Following the method of reservoir partition that
has been applied with Markovian TL treatment,\cite{Shn9815400,Gur9615932,Li05205304}
the TNL--GQME can be readily formulated out as
\begin{align}\label{rhont}
 \dot\rho^{(n)}(t)&=-i{\cal L}\rho^{(n)}(t)-
   \int_0^t\!\mathrm{d}\tau\hat \Sigma^{(0)}(t-\tau) \rho^{(n)}(\tau)
 \nl&\quad
  + \sum_{\sigma=+,-} \int_0^t\!\mathrm{d}\tau
   \hat \Sigma^{(\sigma)}_{\alpha'}(t-\tau)\rho^{(n)}(\tau)
\nl&\quad
  + \sum_{\sigma=+,-} \int_0^t\!\mathrm{d}\tau
   \hat \Sigma^{(\sigma)}_{\alpha}(t-\tau)\rho^{(n+\sigma\cdot 1)}(\tau)
\nl&\quad
 -\delta_{n0}\hat\varrho(t)\, ,
\end{align}
where $\alpha'\neq \alpha$,
with $\alpha$ specifying the junction lead of current counting performed.
The involving kernels, $\hat\Sigma^{(0)}(t)$
and $\hat\Sigma^{(\pm)}_{\alpha}(t)$ [also $\hat\Sigma^{(\pm)}_{\alpha'}(t)$],
had been expressed in terms of
their Laplace frequency transformations in
\Eq{allSig}, followed by the comments on
their associating physical processes.
The inhomogeneous term in \Eq{rhont} is of
\be\label{varrho}
  \hat\varrho(t)=\int^{0}_{-\infty}\!d\tau\hat\Sigma(t-\tau) \rho(\tau).
\ee
It arises as the counting field takes action only after
a given finite time,\cite{Fli08150601}
which is set to be $t=0$ without losing the generality.
Therefore, the temporal argument $t$ in \Eq{rhont}
is nothing but the desired current counting measurement time interval.

  The initial conditions to \Eq{rhont} are
\be\label{rhon0}
  \rho^{(n)}(0)=\delta_{n0}\rho^\textrm{st}, \ \ \text{with \ }
  [i{\cal L}+\Sigma(\omega=0)]\,\rho^{\rm st}=0,
\ee
before counting the number of electrons
passing through the junction.
The steady-state $\rho^\textrm{st}$ can be
evaluated via the second identity in \Eq{rhon0},
which amounts to \Eq{ME0} with $\dot\rho(t)=0$,
together with the normalization ${\rm tr}\rho^\textrm{st}=1$.
 Apparently, the initial conditions
to TNL-GQME contain
the {\em initial system-environment
correlation} via $\rho^\textrm{st}$.
We will see in \Sec{ththeo2}
that $\hat\varrho(t)$ does not
enter directly into the final expression
of noise spectrum. In other words,
the effect of initial system-environment
correlation on the noise spectrum
is dictated by $\rho^\textrm{st}$,
rather than the inhomogeneous component in \Eq{rhont}.

 An alternative approach to GQME (\ref{rhont})
is the introduction of the
counting field $\chi$ at the specified lead ($\alpha$)
of current counting.\cite{Lev93225,Lev964845,Bag03085316}
It results in the modified tunneling Hamiltonian
by $c_{\alpha k}\rightarrow c_{\alpha k} e^{i\chi}$
in \Eq{Hprime0}.
The resulting GQME for the counting field $\chi$--resolved
reduced state, $\rho_{\chi}(t)\equiv \sum_{n} e^{-i n\chi} \rho^{(n)}(t)$, reads
\be\label{ME-1}
 \dot\rho_{\chi}\!(t)=-i{\cal L}\rho_{\chi}\!(t)
  - \int_{0}^t\!\!{\mathrm d}\tau
  \hat \Sigma_{\chi}\!(t-\tau)\rho_{\chi}\!(\tau)
  -\hat\varrho(t)\,,
  \ee
with (noting that $\alpha'\neq \alpha$ the counting lead)
\be\label{chiSigmaw}
  \hat\Sigma_{\chi}\!(t)
= \hat\Sigma^{(0)}(t)
 -\sum_{\sigma=+,-}
   \left[\hat\Sigma^{(\sigma)}_{\alpha'}(t)+
   e^{\sigma i\chi}\hat\Sigma^{(\sigma)}_{\alpha}(t)\right].
\ee
The GQME (\ref{rhont})
can be obtained via the resolution
$\rho^{(n)}(t)=(2\pi)^{-1}\int\!{\mathrm d}\chi\,e^{in\chi} \rho_{\chi}(t)$,
while the conventional QME (\ref{ME0}) that governs
$\rho(t)=\sum_{n}\rho^{(n)}(t)$ is recovered
by setting $\chi=0$, as inferred from
\Eqs{memoryw}--(\ref{allSig}).
Apparently, the initial condition to
\Eq{ME-1} is nothing but
$\rho_{\chi}(t=0)=\rho^\textrm{st}$.
Thus, the temporal argument $t$ in \Eq{ME-1},
the counting-field domain of \Eq{rhont},
does denote the current counting measurement time interval.

\section{Spectrum density of current}
\label{ththeo2}

\subsection{Transport self-energy formalism}
\label{ththeo2A}

 The GQME (\ref{rhont})
is the key dynamics formalism for current counting statistics.
Its Laplace--frequency--domain equivalence for
$\ti\rho^{(n)}(\omega)\equiv \mathscr{L}\big[\rho^{(n)}(t)\big]$
is given by
\begin{align}\label{ME-2}
 &\quad [i({\cal L}-\omega)+\Sigma^{(0)}(\omega)-\Sigma^{(+)}_{\alpha'}(\omega)
 -\Sigma^{(-)}_{\alpha'}(\omega)]\ti\rho^{(n)}(\omega)
\nl&= \Sigma^{(+)}_{\alpha}(\omega)
   \ti\rho^{(n+1)}(\omega)
  + \Sigma^{(-)}_{\alpha}(\omega)
   \ti\rho^{(n-1)}(\omega)
\nl&\quad  + \delta_{n0}[\rho^{\rm st}-\varrho(\omega)]\, .
\end{align}
It will be used directly
in the evaluation of transport current spectrum below.
Note that we have denoted $\alpha$ as the counting lead,
while $\alpha'\neq \alpha$.

 Introduce the transport self-energy functions of
\be\label{calJ_def}
  {\mathcal{J}}^{(\pm)}_\alpha(\omega)
  \equiv \Sigma^{(-)}_\alpha(\omega) \pm \Sigma^{(+)}_\alpha(\omega).
\ee
For the current to the $\alpha$--lead,
$I_{\alpha}(t)=-{\rm Tr}\sum_n n\dot\rho^{(n)}(t)$,
\Eq{ME-2} leads to
$\ti I_\alpha(\omega) \equiv \mathscr{L}[I_{\alpha}(t)]
 = -{\rm Tr}\big[{\mathcal {J}}^{(-)}_\alpha(\omega) \ti\rho(\omega)\big]$.
The stationary current can therefore be evaluated via
\be\label{barI}
  \bar I_\alpha \equiv I^{\rm st}_{\alpha}
=-{\rm Tr}\big[{\mathcal{J}}^{(-)}_\alpha(\omega=0)
   \rho^{\rm st}\big].
\ee
For noise spectrum measurement,
we need also the number density operator, which can be obtained
by using  \Eq{ME-2} as
\begin{align}\label{Nalphaw}
  \ti{N}_{\alpha}(\omega) \equiv \sum_{n} n \ti\rho^{(n)}(\omega)
 =-{\mathcal{G}}(\omega)
 {\mathcal {J}}^{(-)}_\alpha(\omega)
 \rho^{\rm st}/\omega\, ,
\end{align}
where
\be\label{Greenw}
  {\mathcal{G}}(\omega)\equiv[\omega-{\cal L}+i\Sigma(\omega)]^{-1},
\ee
is the  Liouville--space Green's function
for the counting field--free QME (\ref{ME0}).

  Now turn to the shot noise spectrum, defined as
$S(\omega)=\mathscr{F}\big\{\la \delta I(t)\delta I(0)\ra_s\big\}$,
where $\la\delta I(t)\delta I(0)\ra_s$ is the
fluctuating current--current correlation function
that is symmetrized, and
$\mathscr{F}\{\cdots\}$ denotes
the full Fourier transform.
For the total circuit current $I(t)=a I_{\rm L}(t)-b I_{\rm R} (t)$,
the noise spectrum is of
\be \label{sumS}
   S(\omega)=a S_{\rm L}(\omega)+bS_{\rm R}(\omega) -ab\,S_{c}(\omega).
\ee
The involving coefficients that satisfy $a+b=1$ are related to the
symmetry of the junction capacitances.\cite{Bla001}

 For the noise spectrum $S_{\alpha}(\omega)$ at lead $\alpha=$ L or R,
the MacDonald's formula gives directly\cite{Mac62}
\be\label{MacD}
  S_{\alpha}(\omega)=2\omega\!\int_0^{\infty}\!\!{\mathrm d}t\, \sin(\omega t)
  \frac{\mathrm d}{\mathrm dt}[\la n_{\alpha}^2(t) \ra-(\bar I_{\alpha} t)^2].
\ee
Here $\la n^2(t) \ra \equiv \sum_{n} n^2 P(n,t)={\rm Tr}\sum_n n^2 \rho^{(n)}(t)$.
With the help of \Eq{ME-2}, we have
\be\label{dn2t}
   \mathscr{L}\!\left[\frac{\mathrm{d}}{\mathrm{d}t}\la n^2_\alpha(t)\ra\right]
 =2
 {\mathcal {J}}^{(-)}_\alpha(\omega) \ti N_\alpha(\omega)
 + {\mathcal {J}}^{(+)}_\alpha(\omega)\frac{i\rho^{\rm st}}{\omega},
\ee
which together with \Eq{barI} lead to
\begin{align}\label{Salphaw}
  S_{\alpha}(\omega)
&=4\omega\,{\rm Im}
 \big\{ {\rm Tr}[{\mathcal {J}}^{(-)}_\alpha(\omega)
   \ti N_\alpha(\omega)]
 \big\}
\nl&\quad +2\,{\rm Re}
 \big\{ {\rm Tr}[{\mathcal {J}}^{(+)}_\alpha(\omega)
    \rho^{\rm st}]
 \big\}.
\end{align}

 Consider now $S_{c}(\omega)$ in \Eq{sumS}, which is the spectrum
of charge fluctuation $\dot{Q}(t)=-[I_{\rm L}(t)+I_{\rm R}(t)]$
on the central dots.
The current conservation gives
\be\label{Sc}
  S_{c}(\omega) =2S_{\rm LR}(\omega)
 +S_{\rm L}(\omega) +S_{\rm R}(\omega).
\ee
The cross correlation noise spectrum, defined as
$S_{\rm LR}(\omega)={1\over 2}\mathscr{F}\big\{
 \la \delta I_{\rm L}(t)\delta I_{\rm R}(0)\ra_s
+\la \delta I_{\rm R}(t)\delta I_{\rm L}(0)\ra_s\big\}$,
can also be cast to the MacDonald's formula as \cite{Wan04153301,Don05066601}
\[
 S_{\rm LR}(\omega)=2\omega\!\int^\infty_0\!\!{\mathrm d}t \sin(\omega t)
  \frac{{\mathrm d}}{{\mathrm d}t}\!
\left[\la N_{\rm L}(t)N_{\rm R}(t)\ra - (\bar I t)^2\right],
\]
where $\la N_{\rm L}(t)N_{\rm R}(t)\ra={\rm Tr}
\sum_{n_{\rm L}n_{\rm R}}n_{\rm L}n_{\rm R} \rho^{(n_{\rm L}\!,n_{\rm R})}(t)$.
Similarly, with the help of \Eq{ME-2}, we finally obtain
\be\label{S_LR}
  S_{\rm LR}(\omega)
 =2\omega\,{\rm Im}\big\{{\rm Tr}
  [{\mathcal {J}}^{(-)}_{\rm L}(\omega)\ti N_{\rm R}(\omega)
  +{\mathcal {J}}^{(-)}_{\rm R}(\omega)\ti N_{\rm L}(\omega)
  ]\big\}.
\ee
We have thus completed the expression of the dot charge fluctuation
spectrum $S_c(\omega)$. Note that the noise spectrum
may also be formulated by using
the quantum regression theorem,\cite{Agu04206601,Car93}
which however is not applicable to non-Markovian case,
due to the long memory time of the reservoir.
\cite{For96798,Alo05200403}

\subsection{Remarks on non--Markovian shot noise}
\label{ththeo2B}

 The final expressions for evaluating
the spectrum of current fluctuation
comprise therefore \Eqs{Salphaw}--(\ref{S_LR})
together with \Eqs{calJ_def} and (\ref{Nalphaw}).
The key quantities here are ${\cal J}_{\alpha}^{(\pm)}(\omega)$
[\Eq{calJ_def}] or, equivalently,
the transport self--energies $\Sigma_{\alpha}^{(\pm)}(\omega)$
[\Eq{Sigpm}] involved in the
TNL--GQME  in the weak system--reservoir
interaction regime.
Involved in $\Sigma_{\alpha}^{(\pm)}(\omega)$
are the Laplace frequency transformation
of reservoirs correlation functions $C^{(\pm)}_\alpha(t)$,
as defined by \Eq{LapTran}.
The grand fermionic ensemble fluctuation--dissipation
theorem reads\cite{Hau08}
\be\label{FDT}
  C^{(\sigma)}_{\alpha\mu\nu}(t) =
\int^\infty_{-\infty}\!\!{\rm d}\omega\, e^{\sigma i\omega t}
 J_{\alpha\mu\nu}(\omega)f^\sigma_\alpha(\omega)\, .
\ee
Here $f_{\alpha}(\omega)\equiv f^{+}_\alpha(\omega) =
1-f^{-}_\alpha(\omega)$ is the Fermi function
of $\alpha$--electrode;
$J_{\alpha\mu\nu}(\omega)$ denotes the
reservoir spectral density function, which
is diagonal in spin--space; i.e., $J_{\alpha\mu\nu}(\omega)=0$
if the involving system levels $\mu$ and $\nu$ are of different spins.
  Consider the
reservoirs spectral density the form of
$J_{\alpha\mu\nu}(\omega)=J_{\alpha}(\omega)\delta_{\mu\nu}$,
which leads to $C^{(\sigma)}_{\alpha\mu\nu}(\omega)
=C^{(\sigma)}_{\alpha}(\omega)\delta_{\mu\nu}$, where
\be\label{Cspec_disp}
  C^{(\sigma)}_{\alpha}(\omega)\equiv
  \frac{1}{2}[\Delta^{(\pm)}_{\alpha}(\mp\omega)
    + i\Lambda^{(\pm)}_{\alpha}(\mp\omega)].
\ee
with  $\Delta^{(\pm)}_{\alpha}(\omega)
=f^{\pm}_{\alpha}(\omega)J_{\alpha}(\omega)$,
as inferred from \Eq{FDT},
and $\Lambda^{(\pm)}_{\alpha}(\omega)$
are the reservoirs spectrum and dispersion functions,
respectively.
They are related by the Kra mers--Kronig
relation. Without loss physical picture,
we adopt the spectral density, $J_\alpha(\omega)\equiv\sum_k |t_{\alpha k}|^2 \delta(\omega-\epsilon_{\alpha k})$,
the Lorentzian form of
\be\label{Jw_alpha}
 J_{\alpha}(\omega)=\frac{\Gamma_{\alpha} W^2}
  {(\omega-\mu_{\alpha})^2+W^2}.
\ee
Considered here is a rigid homogenous shift in the conduction band
of each electrode by applying the bias voltage,
i.e., $\epsilon_{\alpha k}\rightarrow\epsilon_{\alpha k}+\mu_\alpha$,
so that the occupation of electrons in the leads remains unchanged.
Also, we assume a half-occupied conduction band for each lead, which makes the center of the Lorentzian spectral density coincide with the Fermi level.
We set $\mu_{L/R}=\pm eV/2$, with $\mu^{\rm eq}_\alpha=0$ for each lead at equilibrium.
The Lorentzian $J(\omega)$ leads to the dispersion function analytical expression:
$
  \Lambda^{(\pm)}_{\alpha}(\omega)
= \phi_{\alpha}(\omega) -
 \frac{\Gamma_{\alpha}}{\pi}\left[
  \Psi\Big(\frac{1}{2}+\frac{W}{2\pi k_BT}
 \Big)\pm \pi\frac{\omega-\mu_\alpha}{W}
\right] ,
$
where $\Psi(x)$ is the digamma function, and
\be\label{phi}
 \phi_{\alpha}(\omega) \equiv \frac{\Gamma_\alpha}{\pi}{\rm Re}
\left\{
\Psi\Big(\frac{1}{2}+i\,\frac{\omega-\mu_\alpha}{2\pi k_BT}\Big)
\right\}.
\ee
Physically, the reservoirs spectrum function is related directly
to transfer rate, while the dispersion
function is responsible for energy renormalization.

 Two basic non--Markovian characteristics,
the {\em  finite--frequency--support} and the {\em quasi--step}
features in noise spectrum are anticipated. They arise
from the $f^{\sigma}_{\alpha}(\omega)$ and $J_{\alpha}(\omega)$
components of the integrant in \Eq{FDT}, respectively.
Both components contribute to the
frequency dependence of transport
self--energies $\Sigma^{\pm}_{\alpha}(\omega)$
and thus that of ${\cal J}^{\pm}_{\alpha}(\omega)$
[\Eq{calJ_def}].

 The finite--frequency--support feature arises
from that of $J_{\alpha}(\omega)$,
e.g., \Eq{Jw_alpha} with finite $W$.
The transport self--energies $\Sigma^{(\sigma)}_{\alpha}(\omega)$
or ${\cal J}^{(\sigma)}_{\alpha}(\omega)$ [\Eq{Sigpm}
or (\ref{calJ_def})]
are also of finite frequency support,
approaching to zero when $|\omega-\mu_\alpha|$ goes beyond the bandwidth.
Examine now the expressions of current noise spectrum,
\Eqs{Salphaw} and (\ref{S_LR}). They depend
also the number density operator $\ti N_{\alpha}(\omega)$.
From its definition by \Eq{Nalphaw},
$\ti N_{\alpha}(\omega)$ always approaches to zero as $\omega\rightarrow\infty$.
It leads to $S_{\rm LR}(\omega\rightarrow\infty)\rightarrow 0$
and $S_{\rm L/R}(\omega\rightarrow\infty) \rightarrow 2\,{\rm Re}
 \big\{ {\rm Tr}[{\mathcal {J}}^{(+)}_{\rm L/R}(\omega)
    \rho^{\rm st}]\big\}$.
Therefore the current noise spectrum vanishes as $\omega\rightarrow\infty$,
for any finite bandwidth $W$.
It differs from the case of WBL ($W\rightarrow\infty$),
where $J_{\alpha}(\omega)=\Gamma_{\alpha}$ is constant
and the resulting noise spectrum approaches to a
constant at high frequency limit.

 The quasi--step feature rooted in that of Fermi function
always exists, even in the WBL.
The quasi--step behavior of Fermi function
manifests itself through $C^{(\sigma)}_{\alpha \mu\nu}(\omega)$
to $\Sigma^{\pm}_{\alpha}(\omega)$ or ${\cal J}^{\pm}_{\alpha}(\omega)$
[\Eq{Sigpm} or \Eq{calJ_def}]
in transport current
statistics.\cite{Eng04136602}
As a result, the quasi--step feature in current noise
may reflect the dot energy structure,
including the magnitude of the finite Coulomb interacting $U$,
which will be demonstrated in the coming section.
This characteristic is evident from \Eq{memoryt},
since the frequency involved in
the Fermi function will be replaced by $\omega\pm \epsilon$ or
$\omega\pm (\epsilon+U)$.


 The aforementioned two non--Markovian characteristics,
as just analyzed on the basic of TNL--GQME in the weak
system--reservoirs coupling regime,
will be verified numerically soon.
Note that the TL--GQME and its consequent  Markovian
noise spectrum \cite{Leh02228305,Li05205304,Li05066803,Li07075114,Har06235309}
can be recovered by setting
$\Sigma_{\alpha}^{(\pm)}(\omega)\approx\Sigma_{\alpha}^{(\pm)}(0)$.
Note also that a memory kernel treatment of phonon bath interaction
alone leads to
a frequency dependent total self--energy
$\Sigma(\omega)$ at the conventional QME level,
but retains a frequency independent
$\Sigma_{\alpha}^{(\pm)}$ and is therefore
TL at the GQME level.\cite{Fli08150601}
The TL--GQME that
assumes the frequency--independent $\Sigma^{(\pm)}_{\alpha}$,
even with the inclusion of non-Markovian phonon bath
coupling,\cite{Fli08150601}
will miss the
aforementioned non--Markovian transport characteristics.
The TNL--GQME approach is found to be more suitable
(see \Fig{fig1} below),
as assessed by the exact and nonperturbative
result readily available at least for noninteracting
systems.\cite{Jin10083013}

\section{Numerical demonstrations}
\label{num}

In the following demonstrations, we set $\mu_{\rm L}=-\mu_{\rm R}= eV/2$,
with $\mu^{\rm eq}_{\rm L}=\mu^{\rm eq}_{\rm R}=0$
for the equilibrium electrodes chemical
potentials in the absence of external bias voltage $V$.
We focus on the regime
of $W \gg \Gamma_{\alpha}$,
which is often the case of realistic systems.
This regime would also validate the TNL-GQME to be
a suitable weak system--reservoirs coupling theory,
as supported by our early work on
the spectrum analysis of transient transport current
calculation,\cite{Zhe08093016}
and also by Ref.\,\onlinecite{Zed09} that
reports the exact zero--frequency noise spectrum of
single-resonant-level dot system.
The close comparison with exact results in a simple
system in \Sec{numA} will further favor TNL--GQME while discriminate against
TL--GQME treatment.

 Presented in due course are also
the analytical TNL--GQME results of noise spectrum
in the WBL, together with the approximated Fermi function,
$f_{\alpha}(\omega<\mu_\alpha)\approx 1$ and
$f_{\alpha}(\omega>\mu_\alpha) \approx 0$.
Consequently, the aforementioned
quasi--step non-Markovian characteristic
of noise spectrum will be highlighted analytically by
piecewise functions [cf.\ \Eqs{Fano1} and (\ref{Fano2})].
The finite--frequency--support non-Markovian feature
will appear numerically, for
Lorentzian spectral density with finite bandwidth.

\subsection{Noninteracting single dot}
\label{numA}

\begin{figure}
\centerline{\includegraphics*[width=0.95\columnwidth]{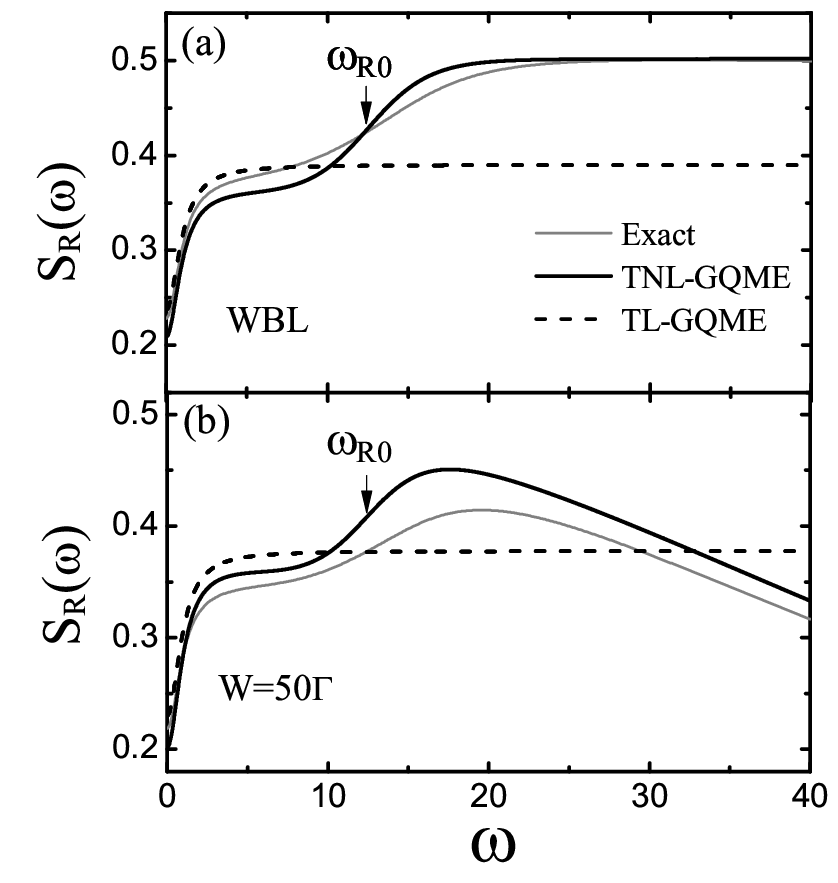}}
\caption{Noise spectrum for right reservoir, evaluated with
three methods: exact (solid gray line), TNL-GQME (solid black
line) and TL-GQME (dashed line) treatments for (a) WBL
and (b) Finite band width of $W=50\Gamma$.
The other parameters are (in arbitrary unit $\Gamma$):
 $\epsilon=5$, $\Gamma_L=\Gamma_R=0.5$, $k_BT=2$, and $eV=15$.}
\label{fig1}
\end{figure}

  Consider first the simple system
of a single spinless level, $H =\epsilon\, d^\dg d$,
as its exact results can be readily
carried out, by using the nonperturbative
GQME theory based on Feynman--Vernon influence
functional.\cite{Jin10083013}
Thus, the demonstrations on this simple system will
not only highlight the aforementioned
two non-Markovian characteristics of shot noise
spectrum, but also more or less justify
the TNL--GQME based noise spectrum calculations in this work.
Figure \ref{fig1} depicts the resulting noise spectra,
evaluated on the basis of the exact (solid in gray), TNL--GQME (solid in black)
and TL--GQME (dashed in black) theories.
Evidently, the TNL--GQME reproduces well, at least qualitatively
all basic features in the entire frequency range,
while the TL--GQME is only applicable
in the low frequency regime of $\omega<\omega_{\alpha0}\equiv|\mu_\alpha-\epsilon|$.
The above observations are supported by
the fact that TNL--GQME is non--Markovian
while TL--GQME is Markovian.
The high frequency regime corresponds to short time scale where
non-Markovian effect is strong, while
the low frequency regime corresponds to long time
scale where non-Markovian effect diminishes.
The non--Markovian quasi--step characteristic
is highlighted in \Fig{fig1}(a) with $W\rightarrow\infty$ (i.e., the WBL),
while the finite--frequency--support feature
is demonstrated in \Fig{fig1}(b) with $W=50\,\Gamma$, 
where $\Gamma\equiv \Gamma_{\rm L}+\Gamma_{\rm R}$.
Apparently, the high--frequency
breakdown of TL--GQME is general, even in the WBL.
The TNL--GQME is the choice of weak system--reservoirs
coupling theory for the entire frequency range.

  The analytical TNL--GQME based results in the WBL
are summarized in the following to highlight the non--Markovian quasi--step feature,
with the approximated Fermi function,
$f_\alpha(\omega)\approx 1$ for $\omega<\mu_\alpha$, and zero otherwise.
Consider also large bias, $\mu_{\rm L}\gg\epsilon\gg\mu_{\rm R}$.
In this case, we can neglect
the dispersion component $\Lambda^{(\pm)}_{\alpha}(\omega)$
of $C_\alpha(\omega)$ in \Eq{Cspec_disp},
as its resultant energy renormalization effect on the present
single--level system is negligible.
The transport current
is $\bar I = \bar I_{\,\mathrm{L}} - \bar I_{\,\mathrm{R}}
=\Gamma_{\rm L}\Gamma_{\rm R}/\Gamma$.
In the low frequency region,
the current noise spectrum calculated by TNL--GQME
is about the same as the
TL--GQME,  due to negligible non-Markovian effect.
The resulting Fano factor reads for $\omega<\omega_{\alpha0}$
\be\label{fano0}
 F_{\alpha}(\omega)\equiv
  \frac{S_{\alpha}(\omega)}{2\bar I}\approx
\frac{\Gamma^2_{\rm L}+\Gamma^2_{\rm R}+\omega^2}
 {\Gamma^2+\omega^2}
= F^{\rm M}_{\alpha}(\omega).
\ee
The last identity is the TL--GQME or Markovian result,
claimed for all frequencies.

 The non-Markovian quasi--step appears
around $\omega = \omega_{\alpha0}$.
The noise spectrum in the WBL behaves then as
$S_{\alpha}(\omega\rightarrow\infty)\rightarrow\Gamma_\alpha$,
leading to the Fano factors asymptotically of
$F_{\rm L}(\omega)\rightarrow(1+\gamma)/2$
and $F_{\rm R}(\omega)\rightarrow(1+\gamma^{-1})/2$.
These results are consistent
with those of Ref.\,\onlinecite{Eng04136602},
which exploited the standard scattering methods exactly.\cite{Bla001,Hau08}
  Apparently, the Fano factors can be larger or smaller than 1,
determined by  the $\gamma\equiv\Gamma_{\rm L}/\Gamma_{\rm R}$ ratio:
$\gamma=1$ assumes the Poissonian noise
for both leads; $\gamma>1$ enhances the noise of the left lead,
while suppresses that of the right; and vice versa.
The reason behind is that
the tunneling rates difference
($\Gamma_{\rm L}\neq\Gamma_{\rm R}$) is like a
dynamical channel blockade, resulting in bunching
and anti-bunching events.
In contrast, the
TL--GQME leads to the Makorvian results
of $F^{\rm M}_{\alpha}(\omega\rightarrow\infty)=1$ for both leads,
regardless the bandwidth.
Actually the finite bandwidth is of
$F_\alpha(\omega\rightarrow\infty)=0$,
as predicted by either TNL--GQME or exact theory;
see Fig.\,\ref{fig1}(b).
This is right the finite-frequency-support characteristics
that restricts the channels
for electron transferring between
the dots and leads accompanied by the energy ($\hbar\omega$)
absorption/emmison of detection.

\subsection{Interacting single dot}
\label{numB}

 The Coulomb interaction case is exemplified with
\be\label{H_ISD}
 H= \epsilon_{\up}\hat n_\up + \epsilon_{\down}\hat n_\down
   +U \hat n_\up \hat n_\down ,
\ee
where $\hat n_\mu=d^\dg_\mu d_\mu$. The annihilation operators are
$d_{\up}=|0\ra\la\up\!\!|+|\!\!\down\ra\la\up\down\!\!|$
and $d_{\down}=|0\ra\la\down\!\!|-|\!\!\up\ra\la\up\down\!\!|$,
with $|0\ra$, $|\!\!\up\ra$, $|\!\!\down\ra$, and $|\!\!\up\down\ra$
denoting the empty, two single--occupation spin states,
and the  double--occupation spin--pair state, respectively,
in the Fock space.
To have the Coulomb interaction effect more transparent,
we set the dot level spin--degenerate,
$\epsilon_\up=\epsilon_\down=\epsilon$,
and focus on the transport in strong Coulomb blockade regime,
$\epsilon+U>\mu_{\rm L}>\epsilon>\mu_{\rm R}$,
where the stationary transport current is
$\bar I=2\Gamma_L\Gamma_R/(2\Gamma_L+\Gamma_R)$.

  For the purpose of comparison later,
we present here the results of TL--GQME based
(Markovian) shot noise spectrum
(denoting $\Gamma_{\rm eff}\equiv 2\Gamma_{\rm L}+\Gamma_{\rm R}$):\cite{Luo07085325}
\be\label{Mark2}
\begin{split}
 F^{\rm M}_{\alpha}(\omega)
= \frac{4\Gamma_{\rm L}^2+\Gamma_{\rm R}^2+\omega^2}
  {\Gamma_{\rm eff}^2+\omega^2},
\ \ \
  F^{\rm M}_{c}(\omega)
= \frac{2\omega^2}{\Gamma_{\rm eff}^2 +\omega^2},
\end{split}
\ee
which assume Poissonian,
$F^{\rm M}_{\rm L}=F^{\rm M}_{\rm R}=1$
and $F^{\rm M}_c=2$, as $\omega\rightarrow\infty$.
We will see soon that the TNL--GQME treatment
will lead to very different behaviors,
due to the increasing non-Markovian effect with increasing
the detection frequency.

\begin{figure}
\centerline{\includegraphics*[width=0.95\columnwidth]{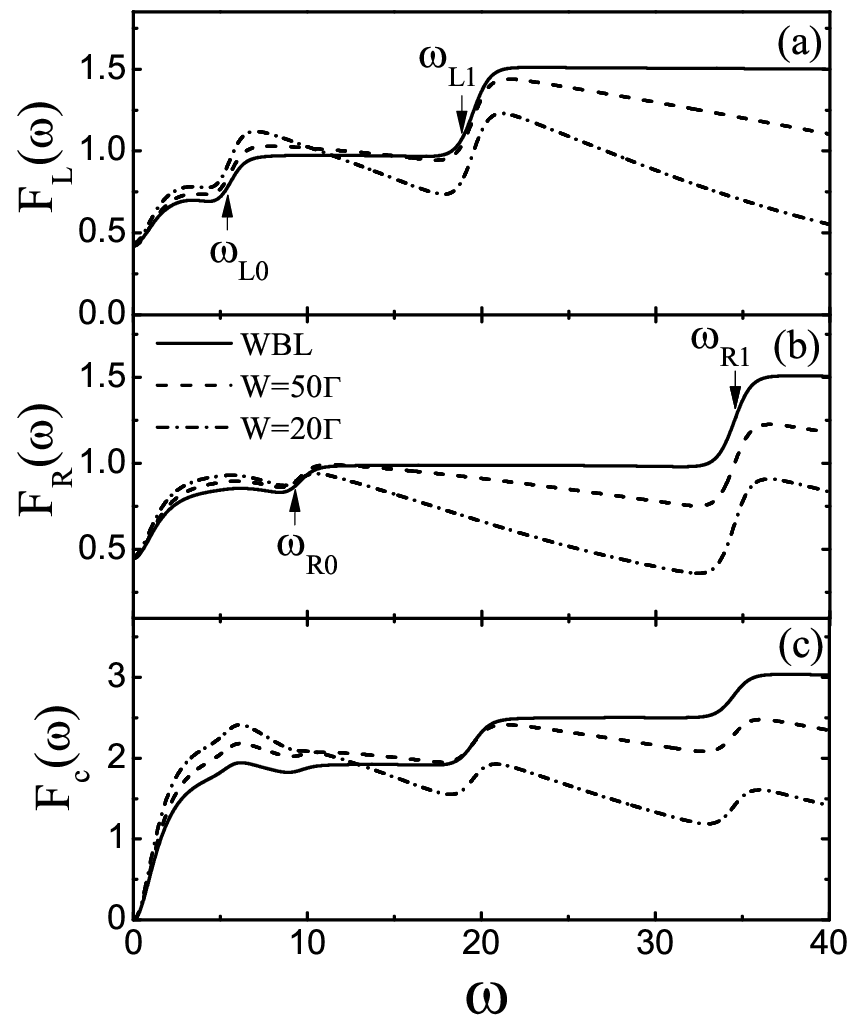}}
\caption{
 Noise spectrum of transport current [in $F(\omega)\equiv S(\omega)/(2\bar I)$]
  through an interacting
 quantum dot, decomposed to
 (a) left junction current, (b) right junction current,
 and (c) charge-number fluctuation
 components, respectively.
 Parameters (in arbitrary unit $\Gamma$) are $U=25$,
 $k_BT=0.5$, and $\epsilon=2$,
 with different values of bandwidth $W$.
 The other parameters are the same as in Fig.\,\ref{fig2}.
\label{fig2}
}
\end{figure}

  Figure \ref{fig2} depicts the noise spectrum of transport current,
with different bandwidths that crossover
the fixed Coulomb interaction parameter $U=25\,\Gamma$
to visualize the interplay between them.
The non-Markovian quasi--step feature
is displayed by a number of quasi--step jumps
around the resonant frequencies of
\be\label{omega01}
 \omega_{\alpha 0}=|\mu_{\alpha}-\epsilon| {\rm\ \  and\ \ }
 \omega_{\alpha 1}=\epsilon+U-\mu_{\alpha},
\ee
with $\alpha=$ L and R. Specifically,
these resonance frequencies (in unit of $\Gamma$) are
$\omega_{{\rm L}0}=5.5$,  $\omega_{{\rm R}0}=9.5$,
$\omega_{{\rm L}1}=19.5$, and $\omega_{{\rm R}1}=34.5$,
for the parameters used in \Fig{fig2}. This characteristic
feature can be used to extract the information of the discrete energy
level of the dot as well as the
magnitude of the Coulomb interaction.
The finite-frequency-support resulted from
finite bandwidth is also apparent
in interacting dots systems.

 To highlight the quasi--steps resonant tunneling characteristics,
we consider the WBL, with the aid of its approximated analytical results
as follows.
Let us start with the region of $\omega>\omega_{\alpha0}$:
\bsube\label{Fano2}
\begin{align}
F_{\alpha}(\omega)
 &=\left\{
 \begin{array}{ll}\frac{1}{2}(1+\gamma_{\alpha})\, ; &\qquad\quad\
  \omega_{\alpha 0}<\omega<\omega_{\alpha 1}
\\
 \frac{1}{2}+\gamma_{\alpha}\, ; &\qquad\quad\
 \omega>\omega_{\alpha 1}
\end{array}\right. ,
\label{Fw2}
 \\
  F_{c}(\omega)&=\left\{ \begin{array}{ll}
  1+\frac{1}{2}(\gamma_{\rm L}+\gamma_{\rm R})\, ; &\
  \omega_{\rm R0}<\omega<\omega_{\rm L1}
  \\
  1+\gamma_{\rm L}+\frac{1}{2}\gamma_{\rm R}\, ; &\
  \omega_{\rm L1}<\omega<\omega_{\rm R1}
    \\
  \frac{1}{2}(3+2\gamma_{\rm L}+\gamma_{\rm R})\, ; &\
  \omega>\omega_{\rm R1}
  \end{array}\right. .
 \end{align}
\esube
Apparently, the noises depending on
$\gamma_{\rm L}\equiv\Gamma_{\rm L}/\Gamma_{\rm R} \equiv 1/\gamma_{\rm R}$
can still be {\it either} super-- {\it or} sub--Poissonian.
Consider for example \Eq{Fw2}, where the two regions
are actually
$\mu_{\rm L} >\epsilon,\epsilon+U >\mu_{\rm R}$
and
$\epsilon+U >\mu_{\rm L} >\epsilon>\mu_{\rm R}$,
representing the weak and the strong Coulomb
interaction cases, respectively.
Evidently the resonant quasi--step characteristics of noise spectrum
are enhanced by Coulomb interaction
[cf.\ \Fig{fig2} or \Eq{Fano2}, with $\omega>\omega_{\alpha 1}$].

\begin{figure}
\centerline{\includegraphics*[width=0.95\columnwidth,angle=0]{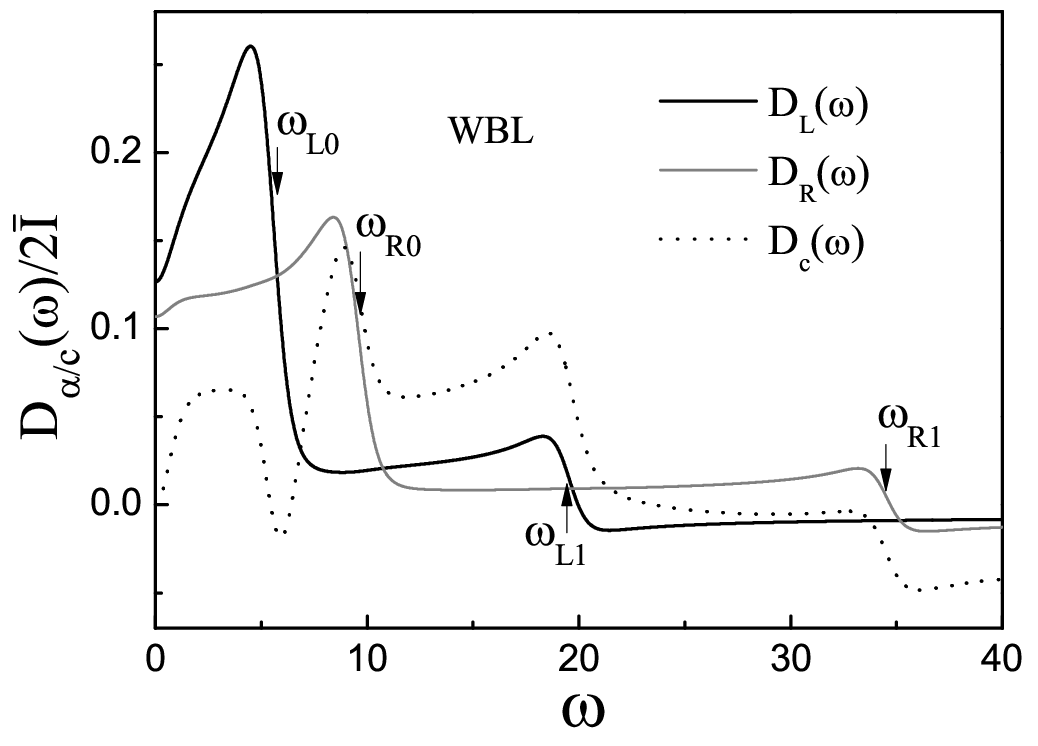}}
\caption{
 The dispersion function contribution to the noise spectrum
  through an interacting
 quantum dot, decomposed to
  left junction current (solid in black), right junction current (solid in gray),
 and charge--number fluctuation (dotted in black)
 components, respectively, with WBL.
 The parameters are the same as in \Fig{fig2}.
\label{figDw}
}
\end{figure}

 Consider the low--frequency region ($\omega<\omega_{\alpha 0}$),
where the noise spectrum can be analyzed via
\be\label{Fano1}
 F_{\alpha/c}(\omega) = F^{\rm M}_{\alpha/c}(\omega)
   -D_{\alpha/c}(\omega)/(2\bar I).
\ee
The first term is just the Markovian result \Eq{Mark2}.
The second term accounts for the renormalization
effect and can be evaluated in the WBL as
\be\label{df}
\begin{split}
D_L(\omega)&=\frac{8\Pi_{\rm L}\Gamma_{\rm L}}
  {\big(\Gamma_{\rm eff}^2+\omega^2\big)\omega} \, , \ \
 D_R(\omega)=\frac{2\Pi'_{\rm R}\Gamma_{\rm R}}
  {\big(\Gamma_{\rm eff}^2+\omega^2\big)\omega} \, ,
  \\
 D_c(\omega)&=
\frac{2(\Phi_{\rm L}+\Phi_{\rm R})
      [\Gamma_{\rm eff}(\Phi_{\rm L}+\Phi_{\rm R})
       -\Gamma_{\rm R}\omega ]
 }{ \Gamma_{\rm eff}^2 +\omega^2  },
  \end{split}
\ee
where
$\Pi'_{\rm R} = \Pi_{\rm R}+\Phi_{\rm R}
\big[(1+\Gamma_{\rm L}/\Gamma_{\rm eff})\omega^2-6\Gamma^2_{\rm L}
 \big]$, and
$\Pi_\alpha =\Phi_{\alpha}[\omega(\Phi_{\rm L}+\Phi_{\rm R})
  +2\Gamma^2_{\alpha'}
  +\omega^2\Gamma_{\alpha'}/\Gamma_{\rm eff}]
  +2 \Gamma_{\rm L}\Gamma_{\rm R}(\Phi_{\rm L}-\Phi_{\rm R})
 $,
with
$\alpha'\neq \alpha$.
These non-Markovian contributions are related
to the dispersion function, through [cf.~\Eq{phi}]
$$ 
 \Lambda^{(\pm)}_{\alpha}(\epsilon-\omega)-\Lambda^{(\pm)}_{\alpha}(\epsilon+\omega)
 \approx \phi_{\alpha}(\epsilon-\omega)
 -\phi_{\alpha}(\epsilon+\omega)
 \equiv   \Phi_\alpha(\omega;\epsilon).
$$ 
The renormalization is important especially
in the low frequency regime, as shown in Fig.\,\ref{figDw}.
Note that $D_c(0)=0$ but
$D_{\alpha}(0)\neq 0$ at zero frequency.
However, the renormalization effect
on the central-dots charge
fluctuation extends wider frequency range
than that on the lead current fluctuation.

\section{Concluding remarks}
\label{sum}

 In summary, we have presented the TNL-GQME
and exploited it in analyzing the
frequency dependence of shot noise spectrum.
The theory itself is the extension of
the conventional TNL-QME \cite{Yan05187}
to quantum measurement problems, via
the standard particle counting $\chi$-field
method.\cite{Lev93225,Lev964845,Bag03085316}
By comparing with the exact results on non-interacting
dots, we numerically demonstrated that the TNL-GQME
is more appropriate than its TL-counterpart.
This observation clearly indicates that the shot noise spectrum
of transport current is generally non-Markovian, even in the WBL
of electron reservoirs leads.

\acknowledgments
Support from GRC Hong Kong (604709),
NNSF China (10904029) and ZJNSF (Y6090345)
 is acknowledged.


\begin{thebibliography}{43}
\expandafter\ifx\csname natexlab\endcsname\relax\def\natexlab#1{#1}\fi
\expandafter\ifx\csname bibnamefont\endcsname\relax
  \def\bibnamefont#1{#1}\fi
\expandafter\ifx\csname bibfnamefont\endcsname\relax
  \def\bibfnamefont#1{#1}\fi
\expandafter\ifx\csname citenamefont\endcsname\relax
  \def\citenamefont#1{#1}\fi
\expandafter\ifx\csname url\endcsname\relax
  \def\url#1{\texttt{#1}}\fi
\expandafter\ifx\csname urlprefix\endcsname\relax\def\urlprefix{URL }\fi
\providecommand{\bibinfo}[2]{#2}
\providecommand{\eprint}[2][]{\url{#2}}

\bibitem[{\citenamefont{Blanter and {B\"{u}ttiker}}(2000)}]{Bla001}
\bibinfo{author}{\bibfnamefont{Y.~M.} \bibnamefont{Blanter}} \bibnamefont{and}
  \bibinfo{author}{\bibfnamefont{M.}~\bibnamefont{{B\"{u}ttiker}}},
  \bibinfo{journal}{Phys. Rep.} \textbf{\bibinfo{volume}{336}},
  \bibinfo{pages}{1} (\bibinfo{year}{2000}).

\bibitem[{Naz(2003)}]{Naz03}
\emph{\bibinfo{title}{Quantum Noise in Mesoscopic Physics}}
  (\bibinfo{publisher}{Kluwer, Dordrecht}, \bibinfo{year}{2003}),
  \bibinfo{note}{edited by Y. V. Nazarov}.

\bibitem[{\citenamefont{Sun and Milburn}(1999)}]{Sun9910748}
\bibinfo{author}{\bibfnamefont{H.~B.} \bibnamefont{Sun}} \bibnamefont{and}
  \bibinfo{author}{\bibfnamefont{G.~J.} \bibnamefont{Milburn}},
  \bibinfo{journal}{Phys. Rev. B} \textbf{\bibinfo{volume}{59}},
  \bibinfo{pages}{10748} (\bibinfo{year}{1999}).

\bibitem[{\citenamefont{Flindt et~al.}(2004)\citenamefont{Flindt,
  {Novotn\'{y}}, and Jauho}}]{Fli04205334}
\bibinfo{author}{\bibfnamefont{C.}~\bibnamefont{Flindt}},
  \bibinfo{author}{\bibfnamefont{T.}~\bibnamefont{{Novotn\'{y}}}},
  \bibnamefont{and} \bibinfo{author}{\bibfnamefont{A.-P.} \bibnamefont{Jauho}},
  \bibinfo{journal}{Phys. Rev. B} \textbf{\bibinfo{volume}{70}},
  \bibinfo{pages}{205334} (\bibinfo{year}{2004}).

\bibitem[{\citenamefont{{Kie{\ss}lich}
  et~al.}(2006)\citenamefont{{Kie{\ss}lich}, Samuelsson, Wacker, and
  {Sch\"{o}ll}}}]{Kie06033312}
\bibinfo{author}{\bibfnamefont{G.}~\bibnamefont{{Kie{\ss}lich}}},
  \bibinfo{author}{\bibfnamefont{P.}~\bibnamefont{Samuelsson}},
  \bibinfo{author}{\bibfnamefont{A.}~\bibnamefont{Wacker}}, \bibnamefont{and}
  \bibinfo{author}{\bibfnamefont{E.}~\bibnamefont{{Sch\"{o}ll}}},
  \bibinfo{journal}{Phys. Rev. B} \textbf{\bibinfo{volume}{73}},
  \bibinfo{pages}{033312} (\bibinfo{year}{2006}).

\bibitem[{\citenamefont{Wang et~al.}(2007)\citenamefont{Wang, Jiao, Li, Li, and
  Yan}}]{Wan07125416}
\bibinfo{author}{\bibfnamefont{S.~K.} \bibnamefont{Wang}},
  \bibinfo{author}{\bibfnamefont{H.}~\bibnamefont{Jiao}},
  \bibinfo{author}{\bibfnamefont{F.}~\bibnamefont{Li}},
  \bibinfo{author}{\bibfnamefont{X.~Q.} \bibnamefont{Li}}, \bibnamefont{and}
  \bibinfo{author}{\bibfnamefont{Y.~J.} \bibnamefont{Yan}},
  \bibinfo{journal}{Phys. Rev. B} \textbf{\bibinfo{volume}{76}},
  \bibinfo{pages}{125416} (\bibinfo{year}{2007}).

\bibitem[{\citenamefont{Braggio et~al.}(2006)\citenamefont{Braggio,
  K{\"{o}nig}, and Fazio}}]{Bra06026805}
\bibinfo{author}{\bibfnamefont{A.}~\bibnamefont{Braggio}},
  \bibinfo{author}{\bibfnamefont{J.}~\bibnamefont{K{\"{o}nig}}},
  \bibnamefont{and} \bibinfo{author}{\bibfnamefont{R.}~\bibnamefont{Fazio}},
  \bibinfo{journal}{Phys. Rev. Lett.} \textbf{\bibinfo{volume}{96}},
  \bibinfo{pages}{026805} (\bibinfo{year}{2006}).

\bibitem[{\citenamefont{Flindt et~al.}(2008)\citenamefont{Flindt, Novotny,
  Braggio, Sassetti, and Jauho}}]{Fli08150601}
\bibinfo{author}{\bibfnamefont{C.}~\bibnamefont{Flindt}},
  \bibinfo{author}{\bibfnamefont{T.}~\bibnamefont{Novotny}},
  \bibinfo{author}{\bibfnamefont{A.}~\bibnamefont{Braggio}},
  \bibinfo{author}{\bibfnamefont{M.}~\bibnamefont{Sassetti}}, \bibnamefont{and}
  \bibinfo{author}{\bibfnamefont{A.-P.} \bibnamefont{Jauho}},
  \bibinfo{journal}{Phys. Rev. Lett.} \textbf{\bibinfo{volume}{100}},
  \bibinfo{pages}{150601} (\bibinfo{year}{2008}).

\bibitem[{\citenamefont{Aguado and Kouwenhoven}(2000)}]{Agu001986}
\bibinfo{author}{\bibfnamefont{R.}~\bibnamefont{Aguado}} \bibnamefont{and}
  \bibinfo{author}{\bibfnamefont{L.~P.} \bibnamefont{Kouwenhoven}},
  \bibinfo{journal}{Phys. Rev. Lett.} \textbf{\bibinfo{volume}{84}},
  \bibinfo{pages}{1986} (\bibinfo{year}{2000}).

\bibitem[{\citenamefont{Zakka-Bajjani et~al.}(2007)\citenamefont{Zakka-Bajjani,
  S\'{e}gala, Portier, Roche, Glattli, Cavanna, and Jin}}]{Zak07236803}
\bibinfo{author}{\bibfnamefont{E.}~\bibnamefont{Zakka-Bajjani}},
  \bibinfo{author}{\bibfnamefont{J.}~\bibnamefont{S\'{e}gala}},
  \bibinfo{author}{\bibfnamefont{F.}~\bibnamefont{Portier}},
  \bibinfo{author}{\bibfnamefont{P.}~\bibnamefont{Roche}},
  \bibinfo{author}{\bibfnamefont{D.~C.} \bibnamefont{Glattli}},
  \bibinfo{author}{\bibfnamefont{A.}~\bibnamefont{Cavanna}}, \bibnamefont{and}
  \bibinfo{author}{\bibfnamefont{Y.}~\bibnamefont{Jin}},
  \bibinfo{journal}{Phys. Rev. Lett.} \textbf{\bibinfo{volume}{99}}
  (\bibinfo{year}{2007}).

\bibitem[{\citenamefont{Onac et~al.}(2006)\citenamefont{Onac, Balestro, van
  Beveren, Hartmann, Nazarov, and Kouwenhoven}}]{Ona06176601}
\bibinfo{author}{\bibfnamefont{E.}~\bibnamefont{Onac}},
  \bibinfo{author}{\bibfnamefont{F.}~\bibnamefont{Balestro}},
  \bibinfo{author}{\bibfnamefont{L.~H.~W.} \bibnamefont{van Beveren}},
  \bibinfo{author}{\bibfnamefont{U.}~\bibnamefont{Hartmann}},
  \bibinfo{author}{\bibfnamefont{Y.~V.} \bibnamefont{Nazarov}},
  \bibnamefont{and} \bibinfo{author}{\bibfnamefont{L.~P.}
  \bibnamefont{Kouwenhoven}}, \bibinfo{journal}{Phys. Rev. Lett.}
  \textbf{\bibinfo{volume}{96}} (\bibinfo{year}{2006}).

\bibitem[{\citenamefont{Aguado and Brandes}(2004)}]{Agu04206601}
\bibinfo{author}{\bibfnamefont{R.}~\bibnamefont{Aguado}} \bibnamefont{and}
  \bibinfo{author}{\bibfnamefont{T.}~\bibnamefont{Brandes}},
  \bibinfo{journal}{Phys. Rev. Lett.} \textbf{\bibinfo{volume}{92}},
  \bibinfo{pages}{206601} (\bibinfo{year}{2004}).

\bibitem[{\citenamefont{Engel and Loss}(2004)}]{Eng04136602}
\bibinfo{author}{\bibfnamefont{H.-A.} \bibnamefont{Engel}} \bibnamefont{and}
  \bibinfo{author}{\bibfnamefont{D.}~\bibnamefont{Loss}},
  \bibinfo{journal}{Phys. Rev. Lett.} \textbf{\bibinfo{volume}{93}},
  \bibinfo{pages}{136602} (\bibinfo{year}{2004}).

\bibitem[{\citenamefont{Entin-Wohlman et~al.}(2007)\citenamefont{Entin-Wohlman,
  Imry, Gurvitz, and Aharony}}]{Ent07193308}
\bibinfo{author}{\bibfnamefont{O.}~\bibnamefont{Entin-Wohlman}},
  \bibinfo{author}{\bibfnamefont{Y.}~\bibnamefont{Imry}},
  \bibinfo{author}{\bibfnamefont{S.~A.} \bibnamefont{Gurvitz}},
  \bibnamefont{and} \bibinfo{author}{\bibfnamefont{A.}~\bibnamefont{Aharony}},
  \bibinfo{journal}{Phys. Rev. B} \textbf{\bibinfo{volume}{75}},
  \bibinfo{pages}{193308} (\bibinfo{year}{2007}).

\bibitem[{\citenamefont{Rothstein et~al.}(2009)\citenamefont{Rothstein,
  Entin-Wohlman, and Aharony}}]{Rot09075307}
\bibinfo{author}{\bibfnamefont{E.~A.} \bibnamefont{Rothstein}},
  \bibinfo{author}{\bibfnamefont{O.}~\bibnamefont{Entin-Wohlman}},
  \bibnamefont{and} \bibinfo{author}{\bibfnamefont{A.}~\bibnamefont{Aharony}},
  \bibinfo{journal}{Phys. Rev. B} \textbf{\bibinfo{volume}{79}},
  \bibinfo{pages}{075307} (\bibinfo{year}{2009}).

\bibitem[{\citenamefont{Yan and Xu}(2005)}]{Yan05187}
\bibinfo{author}{\bibfnamefont{Y.~J.} \bibnamefont{Yan}} \bibnamefont{and}
  \bibinfo{author}{\bibfnamefont{R.~X.} \bibnamefont{Xu}},
  \bibinfo{journal}{Annu. Rev. Phys. Chem.} \textbf{\bibinfo{volume}{56}},
  \bibinfo{pages}{187} (\bibinfo{year}{2005}).

\bibitem[{\citenamefont{Chen et~al.}(2009)\citenamefont{Chen, Zheng, Shi, and
  Yan}}]{Che09094502}
\bibinfo{author}{\bibfnamefont{L.~P.} \bibnamefont{Chen}},
  \bibinfo{author}{\bibfnamefont{R.~H.} \bibnamefont{Zheng}},
  \bibinfo{author}{\bibfnamefont{Q.}~\bibnamefont{Shi}}, \bibnamefont{and}
  \bibinfo{author}{\bibfnamefont{Y.~J.} \bibnamefont{Yan}},
  \bibinfo{journal}{J. Chem. Phys.} \textbf{\bibinfo{volume}{131}},
  \bibinfo{pages}{094502} (\bibinfo{year}{2009}).

\bibitem[{\citenamefont{Lehmann et~al.}(2002)\citenamefont{Lehmann, Kohler,
  {H\"{a}nggi}, and Nitzan}}]{Leh02228305}
\bibinfo{author}{\bibfnamefont{J.}~\bibnamefont{Lehmann}},
  \bibinfo{author}{\bibfnamefont{S.}~\bibnamefont{Kohler}},
  \bibinfo{author}{\bibfnamefont{P.}~\bibnamefont{{H\"{a}nggi}}},
  \bibnamefont{and} \bibinfo{author}{\bibfnamefont{A.}~\bibnamefont{Nitzan}},
  \bibinfo{journal}{Phys. Rev. Lett.} \textbf{\bibinfo{volume}{88}},
  \bibinfo{pages}{228305} (\bibinfo{year}{2002}).

\bibitem[{\citenamefont{Li et~al.}(2005{\natexlab{a}})\citenamefont{Li, Luo,
  Yang, Cui, and Yan}}]{Li05205304}
\bibinfo{author}{\bibfnamefont{X.~Q.} \bibnamefont{Li}},
  \bibinfo{author}{\bibfnamefont{J.~Y.} \bibnamefont{Luo}},
  \bibinfo{author}{\bibfnamefont{Y.~G.} \bibnamefont{Yang}},
  \bibinfo{author}{\bibfnamefont{P.}~\bibnamefont{Cui}}, \bibnamefont{and}
  \bibinfo{author}{\bibfnamefont{Y.~J.} \bibnamefont{Yan}},
  \bibinfo{journal}{Phys. Rev. B} \textbf{\bibinfo{volume}{71}},
  \bibinfo{pages}{205304} (\bibinfo{year}{2005}{\natexlab{a}}).

\bibitem[{\citenamefont{Li et~al.}(2005{\natexlab{b}})\citenamefont{Li, Cui,
  and Yan}}]{Li05066803}
\bibinfo{author}{\bibfnamefont{X.~Q.} \bibnamefont{Li}},
  \bibinfo{author}{\bibfnamefont{P.}~\bibnamefont{Cui}}, \bibnamefont{and}
  \bibinfo{author}{\bibfnamefont{Y.~J.} \bibnamefont{Yan}},
  \bibinfo{journal}{Phys. Rev. Lett.} \textbf{\bibinfo{volume}{94}},
  \bibinfo{pages}{066803} (\bibinfo{year}{2005}{\natexlab{b}}).

\bibitem[{\citenamefont{Li and Yan}(2007)}]{Li07075114}
\bibinfo{author}{\bibfnamefont{X.~Q.} \bibnamefont{Li}} \bibnamefont{and}
  \bibinfo{author}{\bibfnamefont{Y.~J.} \bibnamefont{Yan}},
  \bibinfo{journal}{Phys. Rev. B} \textbf{\bibinfo{volume}{75}},
  \bibinfo{pages}{075114} (\bibinfo{year}{2007}).

\bibitem[{\citenamefont{Harbola et~al.}(2006)\citenamefont{Harbola, Esposito,
  and Mukamel}}]{Har06235309}
\bibinfo{author}{\bibfnamefont{U.}~\bibnamefont{Harbola}},
  \bibinfo{author}{\bibfnamefont{M.}~\bibnamefont{Esposito}}, \bibnamefont{and}
  \bibinfo{author}{\bibfnamefont{S.}~\bibnamefont{Mukamel}},
  \bibinfo{journal}{Phys. Rev. B} \textbf{\bibinfo{volume}{74}},
  \bibinfo{pages}{235309} (\bibinfo{year}{2006}).

\bibitem[{\citenamefont{Jin et~al.}(2010)\citenamefont{Jin, Tu, Zhang, and
  Yan}}]{Jin10083013}
\bibinfo{author}{\bibfnamefont{J.~S.} \bibnamefont{Jin}},
  \bibinfo{author}{\bibfnamefont{M.~W.-Y.} \bibnamefont{Tu}},
  \bibinfo{author}{\bibfnamefont{W.-M.} \bibnamefont{Zhang}}, \bibnamefont{and}
  \bibinfo{author}{\bibfnamefont{Y.~J.} \bibnamefont{Yan}},
  \bibinfo{journal}{New J. Phys.} \textbf{\bibinfo{volume}{12}},
  \bibinfo{pages}{083013} (\bibinfo{year}{2010}).

\bibitem[{\citenamefont{Jin et~al.}(2008)\citenamefont{Jin, Zheng, and
  Yan}}]{Jin08234703}
\bibinfo{author}{\bibfnamefont{J.~S.} \bibnamefont{Jin}},
  \bibinfo{author}{\bibfnamefont{X.}~\bibnamefont{Zheng}}, \bibnamefont{and}
  \bibinfo{author}{\bibfnamefont{Y.~J.} \bibnamefont{Yan}},
  \bibinfo{journal}{J. Chem. Phys.} \textbf{\bibinfo{volume}{128}},
  \bibinfo{pages}{234703} (\bibinfo{year}{2008}).

\bibitem[{\citenamefont{Zheng et~al.}(2008{\natexlab{a}})\citenamefont{Zheng,
  Jin, and Yan}}]{Zhe08184112}
\bibinfo{author}{\bibfnamefont{X.}~\bibnamefont{Zheng}},
  \bibinfo{author}{\bibfnamefont{J.~S.} \bibnamefont{Jin}}, \bibnamefont{and}
  \bibinfo{author}{\bibfnamefont{Y.~J.} \bibnamefont{Yan}},
  \bibinfo{journal}{J. Chem. Phys.} \textbf{\bibinfo{volume}{129}},
  \bibinfo{pages}{184112} (\bibinfo{year}{2008}{\natexlab{a}}).

\bibitem[{\citenamefont{Zheng et~al.}(2008{\natexlab{b}})\citenamefont{Zheng,
  Jin, and Yan}}]{Zhe08093016}
\bibinfo{author}{\bibfnamefont{X.}~\bibnamefont{Zheng}},
  \bibinfo{author}{\bibfnamefont{J.~S.} \bibnamefont{Jin}}, \bibnamefont{and}
  \bibinfo{author}{\bibfnamefont{Y.~J.} \bibnamefont{Yan}},
  \bibinfo{journal}{New J. Phys.} \textbf{\bibinfo{volume}{10}},
  \bibinfo{pages}{093016} (\bibinfo{year}{2008}{\natexlab{b}}).

\bibitem[{\citenamefont{Levitov and Lesovik}(1993)}]{Lev93225}
\bibinfo{author}{\bibfnamefont{L.~S.} \bibnamefont{Levitov}} \bibnamefont{and}
  \bibinfo{author}{\bibfnamefont{G.~B.} \bibnamefont{Lesovik}},
  \bibinfo{journal}{Pis'ma Zh. Eksp. Teor. Fiz.} \textbf{\bibinfo{volume}{58}},
  \bibinfo{pages}{225} (\bibinfo{year}{1993}).

\bibitem[{\citenamefont{Levitov et~al.}(1996)\citenamefont{Levitov, Lee, and
  Lesovik}}]{Lev964845}
\bibinfo{author}{\bibfnamefont{L.~S.} \bibnamefont{Levitov}},
  \bibinfo{author}{\bibfnamefont{H.~W.} \bibnamefont{Lee}}, \bibnamefont{and}
  \bibinfo{author}{\bibfnamefont{G.~B.} \bibnamefont{Lesovik}},
  \bibinfo{journal}{J. Math. Phys.} \textbf{\bibinfo{volume}{37}},
  \bibinfo{pages}{4845} (\bibinfo{year}{1996}).

\bibitem[{\citenamefont{Bagrets and Nazarov}(2003)}]{Bag03085316}
\bibinfo{author}{\bibfnamefont{D.~A.} \bibnamefont{Bagrets}} \bibnamefont{and}
  \bibinfo{author}{\bibfnamefont{Y.~V.} \bibnamefont{Nazarov}},
  \bibinfo{journal}{Phys. Rev. B} \textbf{\bibinfo{volume}{67}},
  \bibinfo{pages}{085316} (\bibinfo{year}{2003}).

\bibitem[{\citenamefont{Yan}(1998)}]{Yan982721}
\bibinfo{author}{\bibfnamefont{Y.~J.} \bibnamefont{Yan}},
  \bibinfo{journal}{Phys. Rev. A} \textbf{\bibinfo{volume}{58}},
  \bibinfo{pages}{2721} (\bibinfo{year}{1998}).

\bibitem[{\citenamefont{Makhlin et~al.}(2001)\citenamefont{Makhlin,
  {Sch\"{o}n}, and Shnirman}}]{Mak01357}
\bibinfo{author}{\bibfnamefont{Y.}~\bibnamefont{Makhlin}},
  \bibinfo{author}{\bibfnamefont{G.}~\bibnamefont{{Sch\"{o}n}}},
  \bibnamefont{and} \bibinfo{author}{\bibfnamefont{A.}~\bibnamefont{Shnirman}},
  \bibinfo{journal}{Rev. Mod. Phys.} \textbf{\bibinfo{volume}{73}},
  \bibinfo{pages}{357} (\bibinfo{year}{2001}).

\bibitem[{\citenamefont{Shnirman et~al.}(2002)\citenamefont{Shnirman, Mozyrsky,
  and Martin}}]{Shn0211618}
\bibinfo{author}{\bibfnamefont{A.}~\bibnamefont{Shnirman}},
  \bibinfo{author}{\bibfnamefont{D.}~\bibnamefont{Mozyrsky}}, \bibnamefont{and}
  \bibinfo{author}{\bibfnamefont{I.}~\bibnamefont{Martin}},
  \bibinfo{journal}{LANL e-print cond-mat/0211618}  (\bibinfo{year}{2002}).

\bibitem[{\citenamefont{Shnirman and {Sch\"{o}n}}(1998)}]{Shn9815400}
\bibinfo{author}{\bibfnamefont{A.}~\bibnamefont{Shnirman}} \bibnamefont{and}
  \bibinfo{author}{\bibfnamefont{G.}~\bibnamefont{{Sch\"{o}n}}},
  \bibinfo{journal}{Phys. Rev. B} \textbf{\bibinfo{volume}{57}},
  \bibinfo{pages}{15400} (\bibinfo{year}{1998}).

\bibitem[{\citenamefont{Gurvitz and Prager}(1996)}]{Gur9615932}
\bibinfo{author}{\bibfnamefont{S.~A.} \bibnamefont{Gurvitz}} \bibnamefont{and}
  \bibinfo{author}{\bibfnamefont{Y.~S.} \bibnamefont{Prager}},
  \bibinfo{journal}{Phys. Rev. B} \textbf{\bibinfo{volume}{53}},
  \bibinfo{pages}{15932} (\bibinfo{year}{1996}).

\bibitem[{\citenamefont{MacDonald}(1962)}]{Mac62}
\bibinfo{author}{\bibfnamefont{D.~K.~C.} \bibnamefont{MacDonald}},
  \emph{\bibinfo{title}{Noise and Fluctuations: An Introduction}}
  (\bibinfo{publisher}{Wiley}, \bibinfo{address}{New York},
  \bibinfo{year}{1962}), \bibinfo{note}{ch.\ 2.2.1}.

\bibitem[{\citenamefont{Wang et~al.}(2004)\citenamefont{Wang, Wang, and
  Guo}}]{Wan04153301}
\bibinfo{author}{\bibfnamefont{B.}~\bibnamefont{Wang}},
  \bibinfo{author}{\bibfnamefont{J.}~\bibnamefont{Wang}}, \bibnamefont{and}
  \bibinfo{author}{\bibfnamefont{H.}~\bibnamefont{Guo}},
  \bibinfo{journal}{Phys. Rev. B} \textbf{\bibinfo{volume}{69}},
  \bibinfo{pages}{153301} (\bibinfo{year}{2004}).

\bibitem[{\citenamefont{Dong et~al.}(2005)\citenamefont{Dong, Cui, and
  Lei}}]{Don05066601}
\bibinfo{author}{\bibfnamefont{B.}~\bibnamefont{Dong}},
  \bibinfo{author}{\bibfnamefont{H.~L.} \bibnamefont{Cui}}, \bibnamefont{and}
  \bibinfo{author}{\bibfnamefont{X.~L.} \bibnamefont{Lei}},
  \bibinfo{journal}{Phys. Rev. Lett.} \textbf{\bibinfo{volume}{94}},
  \bibinfo{pages}{066601} (\bibinfo{year}{2005}).

\bibitem[{\citenamefont{Carmichael}(1993)}]{Car93}
\bibinfo{author}{\bibfnamefont{H.~J.} \bibnamefont{Carmichael}},
  \emph{\bibinfo{title}{An Open System Approach to Quantum Optics}}
  (\bibinfo{publisher}{Spring-Verlag}, \bibinfo{address}{Berlin},
  \bibinfo{year}{1993}).

\bibitem[{\citenamefont{Ford and {O'Connell}}(1996)}]{For96798}
\bibinfo{author}{\bibfnamefont{G.~W.} \bibnamefont{Ford}} \bibnamefont{and}
  \bibinfo{author}{\bibfnamefont{R.~F.} \bibnamefont{{O'Connell}}},
  \bibinfo{journal}{Phys. Rev. Lett.} \textbf{\bibinfo{volume}{77}},
  \bibinfo{pages}{798} (\bibinfo{year}{1996}).

\bibitem[{\citenamefont{Alonso and {de Vega}}(2005)}]{Alo05200403}
\bibinfo{author}{\bibfnamefont{D.}~\bibnamefont{Alonso}} \bibnamefont{and}
  \bibinfo{author}{\bibfnamefont{I.}~\bibnamefont{{de Vega}}},
  \bibinfo{journal}{Phys. Rev. Lett.} \textbf{\bibinfo{volume}{94}},
  \bibinfo{pages}{200403} (\bibinfo{year}{2005}).

\bibitem[{\citenamefont{Haug and Jauho}(2008)}]{Hau08}
\bibinfo{author}{\bibfnamefont{H.}~\bibnamefont{Haug}} \bibnamefont{and}
  \bibinfo{author}{\bibfnamefont{A.-P.} \bibnamefont{Jauho}},
  \emph{\bibinfo{title}{Quantum Kinetics in Transport and Optics of
  Semiconductors}} (\bibinfo{publisher}{Springer-Verlag},
  \bibinfo{address}{Berlin}, \bibinfo{year}{2008}), \bibinfo{edition}{2nd} ed.,
  \bibinfo{note}{springer Series in Solid-State Sciences 123}.

\bibitem[{\citenamefont{Zedler et~al.}(2009)\citenamefont{Zedler, Schaller,
  Kiesslich, Emary, and Brandes}}]{Zed09}
\bibinfo{author}{\bibfnamefont{P.}~\bibnamefont{Zedler}},
  \bibinfo{author}{\bibfnamefont{G.}~\bibnamefont{Schaller}},
  \bibinfo{author}{\bibfnamefont{G.}~\bibnamefont{Kiesslich}},
  \bibinfo{author}{\bibfnamefont{C.}~\bibnamefont{Emary}}, \bibnamefont{and}
  \bibinfo{author}{\bibfnamefont{T.}~\bibnamefont{Brandes}},
  \bibinfo{journal}{arXiv:0902.2118v1}  (\bibinfo{year}{2009}).

\bibitem[{\citenamefont{Luo et~al.}(2007)\citenamefont{Luo, Li, and
  Yan}}]{Luo07085325}
\bibinfo{author}{\bibfnamefont{J.~Y.} \bibnamefont{Luo}},
  \bibinfo{author}{\bibfnamefont{X.~Q.} \bibnamefont{Li}}, \bibnamefont{and}
  \bibinfo{author}{\bibfnamefont{Y.~J.} \bibnamefont{Yan}},
  \bibinfo{journal}{Phys. Rev. B} \textbf{\bibinfo{volume}{76}},
  \bibinfo{pages}{085325} (\bibinfo{year}{2007}).

\end{thebibliography}
\end{document}